\documentclass[a4paper]{jpconf}
\usepackage{graphicx}
\usepackage{cite}
\usepackage{verbatim}
\usepackage{colortbl}
\usepackage{pifont}
\usepackage{amsmath,amsfonts,amssymb}
\usepackage{feynmp}

\newcommand{\gm}{\gamma^\mu}
\newcommand{\smn}{\sigma^{\mu \nu}}
\newcommand{\Wmn}{W_{\mu \nu}}
\newcommand{\Bmn}{B_{\mu \nu}}

\newcommand{\dvz}{d_V^Z}
\newcommand{\daz}{d_A^Z}
\newcommand{\dva}{d_V^\gamma}
\newcommand{\daa}{d_A^\gamma}

\newcommand{\RE}{\text{Re}}
\newcommand{\IM}{\text{Im}}

\begin{document}

\title{Study of Top Effective Operators at the ILC}

\author{Miguel C. N. Fiolhais}

\address{LIP, Department of Physics, University of Coimbra, 3004-516 Coimbra, Portugal}

\ead{miguel.fiolhais@cern.ch}

\begin{abstract}

In this paper we study the effect of new physics contributions to 
the top quark pair production ($t\bar{t}$) in a possible future linear collider, such as the International Linear Collider (ILC).
The use of a dimension-six gauge invariant effective operator approach allows to compare the prospected results 
at the ILC with the current ones obtained at the Large Hadron Collider (LHC), both in neutral and charged current processes.
We also prove that the use of specific observables, together with a combination of measurements in different polarized 
beam scenarios and with different center-of-mass energies, allows to disentangle different effective operator 
contributions and significantly improve the limits on the anomalous couplings with respect to the LHC.

\end{abstract}

\section{Introduction}

The ILC is a possible future linear collider, with an estimated length of 30-50 km to accelerate and collide electrons 
and positrons at a center-of-mass energy of 500 GeV.
Since collisions between electrons and positrons are much easier to analyze than hadronic collisions, the ILC is expected 
to deliver several precision measurements, which provide interesting tests to the Standard Model (SM) predictions.

The top quark is the most massive elementary particle discovered to date, and therefore, a natural candidate for 
the search of new physics beyond the SM. The precision measurement of its properties, 
in particular its couplings, may provide useful insights on possible new physics contributions at a higher energy scale.
The realization of these precision measurements of the top quark properties in a possible future ILC would, therefore, 
provide an interesting complement to the direct searches being carried out at the LHC.

The effect of new physics contributions to the top quark interactions can be
parameterized in a model-independent form, above the electroweak symmetry breaking scale, in terms of 
gauge invariant dimension-six effective operators~\cite{Buchmuller:1985jz},
\begin{equation}
\mathcal{L}_\text{eff} = \sum \frac{C_x}{\Lambda^2} O_x + \dots \,,
\label{ec:effL}
\end{equation}
where $O_x$ are dimension-six effective operators, invariant under the SM gauge
symmetry $\mathrm{SU}(3)_c \times \mathrm{SU}(2)_L \times \mathrm{U}(1)_Y$, characterized by the dimensionless constants $C_x$ and a new physics scale $\Lambda$.
Unlike previous approaches \cite{Grzadkowski:1996kn,Grzadkowski:1998bh,Grzadkowski:1999iq,Grzadkowski:2000nx,Abe:2001nqa,Devetak:2010na}, the effective operator framework 
allows to reduce the number of independent parameters entering fermion trilinear interactions \cite{AguilarSaavedra:2008zc,AguilarSaavedra:2009mx}, and 
grants direct comparisons between measurements of different top quark vertices, such as the $Wtb$ and $Zt\bar{t}$ vertices, in different colliders.

\begin{figure}[!t]
\begin{center}
\vspace{-3.cm}
\includegraphics[height=12.75cm]{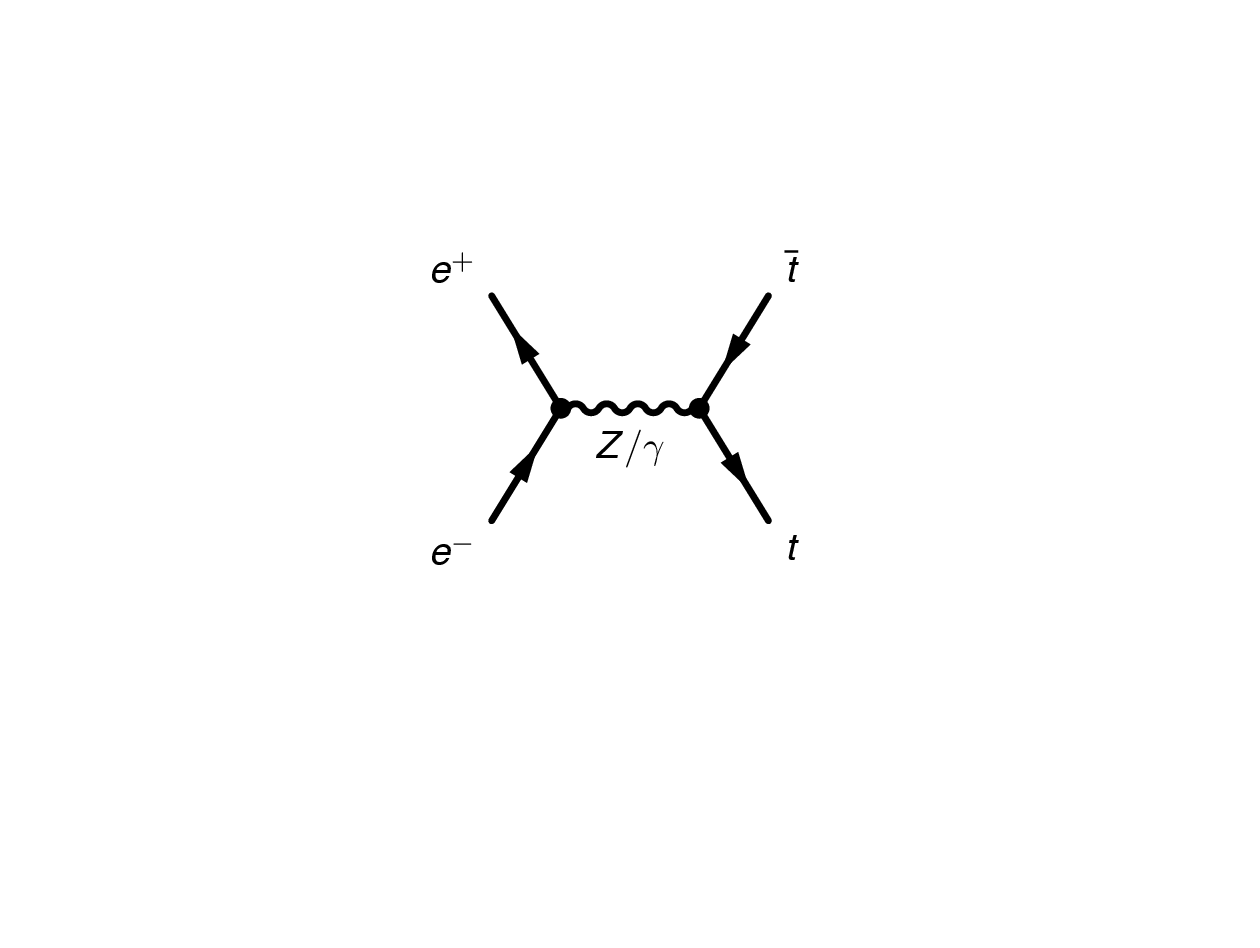}
\vspace{-5.5cm}
\caption{Feynman diagram of the top quark pair production at the ILC through the s-channel.}
\label{figure:eett}
\end{center}
\end{figure}

Among the dimension-six gauge invariant effective operators, there are only five non-redudant operators which contribute to the $t\bar{t}$ production at the ILC 
(depicted in Figure~\ref{figure:eett}), or in other words, to the $Zt\bar{t}$ and $\gamma t\bar{t}$ 
interaction vertices \cite{AguilarSaavedra:2008zc,AguilarSaavedra:2009mx,AguilarSaavedra:2012vh}:

\begin{align}
& O_{\phi q}^{(3,3+3)} = i \left[ \phi^\dagger (\tau^I D_\mu - \overleftarrow
 D_\mu \tau^I) \phi \right] \;
       (\bar q_{L3} \gm \tau^I q_{L3}) \,,
  && O_{uW}^{33} = (\bar q_{L3} \smn \tau^I t_{R}) \tilde \phi \, \Wmn^I \,, \notag \\
& O_{\phi q}^{(1,3+3)} = i (\phi^\dagger \overleftrightarrow D_\mu \phi) (\bar q_{L3} \gm q_{L3}) \,,
  && O_{uB\phi}^{33} = (\bar q_{L3} \smn t_{R}) \tilde \phi \, \Bmn \,, \notag \\
& O_{\phi u}^{3+3} = i (\phi^\dagger \overleftrightarrow D_\mu \phi) (\bar t_{R} \gm t_{R}) \,,
\label{ec:Oall}
\end{align}
using standard notation where $\tau^I$ are the Pauli matrices, $q_{L3}$ is the left-handed third generation quark doublet, $t_R$ is the right-handed top quark singlet, $\phi$ is the SM 
Higgs doublet, $\tilde \phi = i \tau^2 \phi^*$, $\Wmn^I$ and $\Bmn$ is the $\text{SU}(2)_L$ and $\text{U}(1)_Y$ field strength tensors, 
respectively, $D_\mu$ ($\overleftarrow D_\mu$) the covariant derivative acting on the right (left) and $\overleftrightarrow D_\mu = D_\mu - \overleftarrow D_\mu$. 
Since the three operators in the left column of equation~(\ref{ec:Oall}) are Hermitian, their coefficients must be real.

After the spontaneous symmetry breaking, the most general dimension-six $Zt\bar{t}$ and $\gamma t\bar{t}$ lagrangians read
\begin{eqnarray}
\mathcal{L}_{Ztt} & = & - \frac{g}{2 c_W} \bar t \, \gm \left( c_L^t P_L
  + c_R^t P_R \right) t\; Z_\mu  - \frac{g}{2 c_W} \bar t \, \frac{i \smn q_\nu}{M_Z}
  \left( \dvz + i \daz \gamma_5 \right) t\; Z_\mu  \,, \\
\mathcal{L}_{\gamma tt} & = & - e Q_t \bar t \, \gm  t\; A_\mu  - e \bar t \, \frac{i \smn q_\nu}{m_t} \left( \dva + i \daa \gamma_5 \right) t\; A_\mu  \,,
\label{ec:Ztt}
\end{eqnarray}
with $c_L^t = X_{tt}^L - 2 s_W^2 Q_t$, $c_R^t = X_{tt}^R - 2 s_W^2 Q_t$ ($Q_t=2/3$ is the top quark electric charge) and
\begin{align}
& X_{tt}^L = 1  + \left[  C_{\phi q}^{(3,3+3)} - C_{\phi q}^{(1,3+3)}
\right]  \frac{v^2}{\Lambda^2} \,,
&& \dvz = {\sqrt 2} \, \RE \left[ c_W  C_{uW}^{33} - s_W C_{uB\phi}^{33} \right]  \frac{v^2}{\Lambda^2}
\,, \notag \\
& X_{tt}^R =  - C_{\phi u}^{3+3} \frac{v^2}{\Lambda^2} \,,
&& \daz = {\sqrt 2} \, \IM \left[ c_W C_{uW}^{33} - s_W  C_{uB\phi}^{33} \right] \frac{v^2}{\Lambda^2}
\,,
\label{ec:Ztt2}
\end{align}
and,
\begin{eqnarray}
\dva & = & \frac{\sqrt 2}{e} \, \RE \left[s_W  C_{uW}^{33} + c_W C_{uB\phi}^{33} \right] \frac{v m_t}{\Lambda^2}
  \,, \notag \\
\daa & = & \frac{\sqrt 2}{e} \, \IM \left[s_W  C_{uW}^{33} +  c_W C_{uB\phi}^{33} \right] \frac{v m_t}{\Lambda^2} \,.
\label{ec:gatt2}
\end{eqnarray}

Since the SM bottom quark couplings have already been probed with great precision at PETRA, LEP and SLD, it is reasonable 
to assume the following approximation~\cite{AguilarSaavedra:2008zc,AguilarSaavedra:2009mx,AguilarSaavedra:2012vh,Aguilar-Saavedra:2013pxa}: 
\begin{equation}
C_{\phi q}^{(1,3+3)} \simeq -  C_{\phi q}^{(3,3+3)} \,,
\label{ec:Crel}
\end{equation}
to cancel out the non-SM contributions to the $Z b_L b_L$ vertex. The same equality also appears in other SM extensions, such as in new charge $2/3$ singlets~\cite{delAguila:1998tp,AguilarSaavedra:2002kr,delAguila:2000aa,delAguila:2000rc}.

\section{Direct comparison between ILC and LHC}

\begin{figure}[ht]
\begin{center}
\begin{tabular}{cc}
\includegraphics[height=4.75cm]{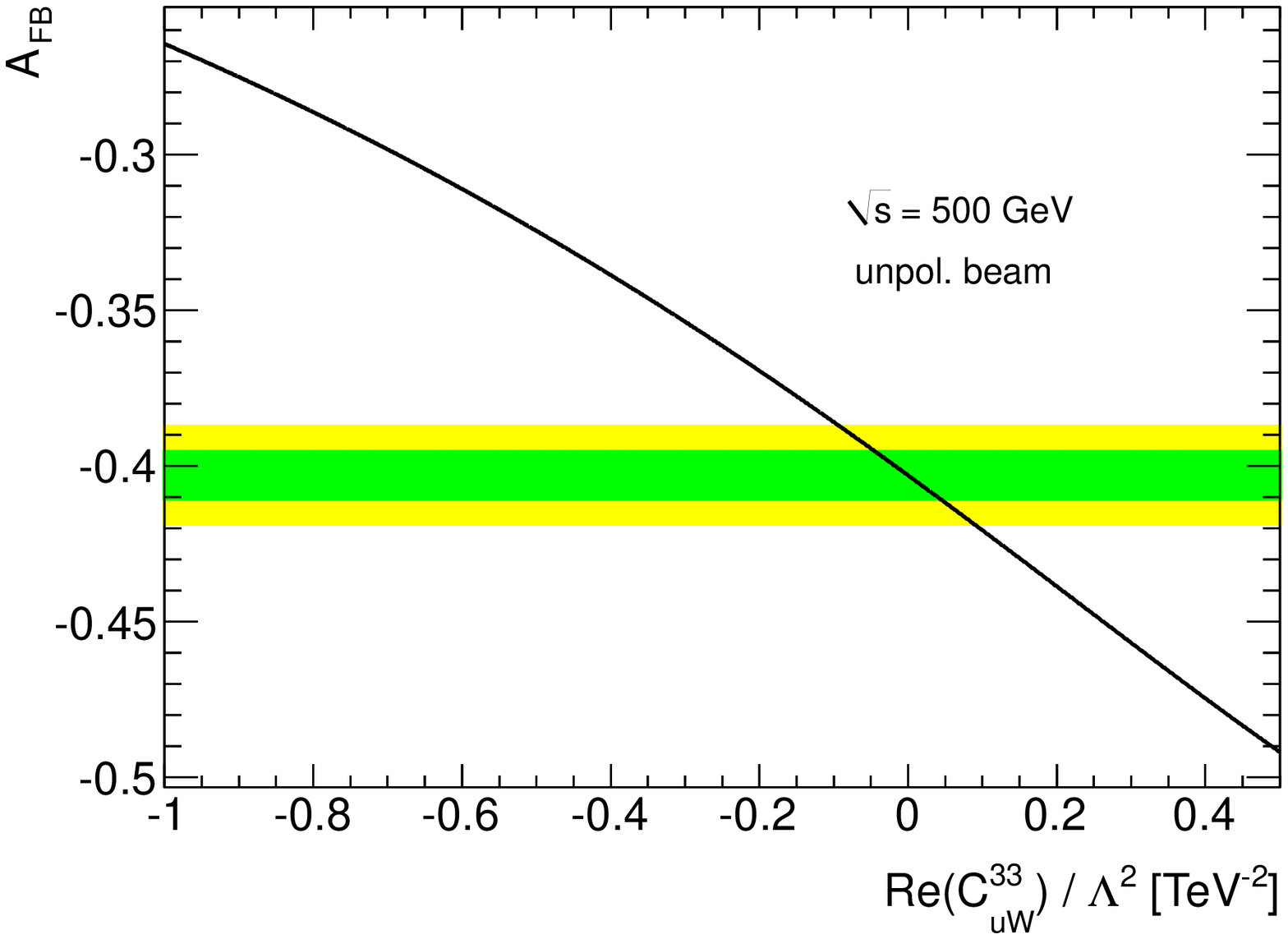} & \includegraphics[height=5.cm]{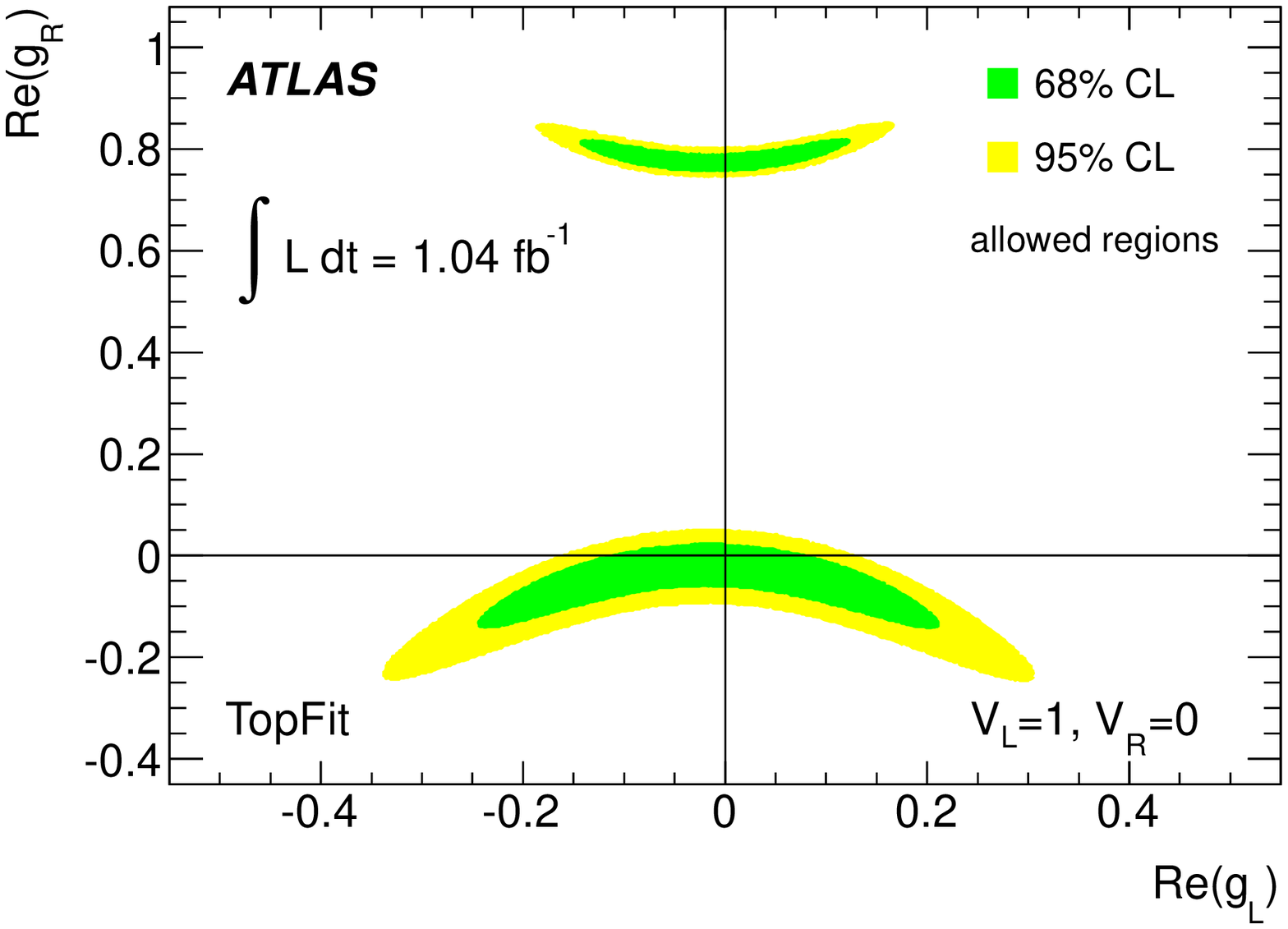}
\end{tabular}
\caption{Left: dependence of the FB asymmetry on $\RE\,C_{uW}^{33}$. Right: allowed regions at 68\% and
    95\% confidence level (CL) for the $Wtb$ anomalous couplings (extracted from \cite{Aad:2012ky}).}
\label{fig:sigma2}
\end{center}
\end{figure}

In order to probe the effective operator coefficients at the ILC, specific observables were used, 
such as the total cross-sections and forward-backward asymmetries. 
The strong dependency of these observables with the effective operators coefficients enhances the possibility to further constrain the anomalous contributions, and 
allows to disentangle different operators contributions\footnote{The quadratic terms were kept in the operator coefficients,
which is consistent with the $1/\Lambda^2$ expansion of the effective operator framework~\cite{AguilarSaavedra:2010sq}.}.
Furthermore, since there are two effective operators, $O_{\phi q}^{(3,3+3)}$ and $O_{uW}^{33}$, which modify both the $Zt\bar{t}$, $\gamma t\bar{t}$ and $Wtb$ vertices,
the prospected results at the ILC can be compared with the current ones at the LHC.
For example, in Figure~\ref{fig:sigma2} (left), the dependency of an ILC observable, the unpolarized FB asymmetry, is presented 
for $\RE \, C_{uW}^{33}$, within the window range of the current limits extracted by the ATLAS Collaboration~\cite{Aad:2012ky}, in Figure~\ref{fig:sigma2} (right), 
through the measurement of the $W$ boson helicity fractions in top quark decays\footnote{Note that $g_{\rm R} = \sqrt 2 C_{uW}^{33} \nu^2 / \Lambda^2$.}.
On the left plot, the yellow and green bands around the SM value correspond to $1\sigma$ and $2\sigma$ variations, respectively,
for total total uncertainties of 5\% in the cross section and 2\% in the asymmetry~\cite{Doublet:2012wf}.
These results show the great potential of the ILC in improving the limits on the effective operator coefficients with respect to the LHC, 
due to smaller uncertainties and enhanced dependencies. As an additional exercise, 
the sensitivity to the new physics scale $\Lambda$ can be extended up to 4.5 TeV for an operator coefficient equal to the unity.
The anti-Hermitian part of this operator have also been probed at the ATLAS experiment~\cite{ATLAS-CONF-2013-032}, 
with a CP-violating asymmetry $A_\text{FB}^N$, defined for polarized top decays~\cite{AguilarSaavedra:2010nx}.
Even though it does not interfere with the SM in CP-conserving observables, such as the total cross-section, 
the sensitivity at the ILC is similar to the one at the LHC.

For other operators, such as $O_{\phi q}^{(3,3+3)}$ and $O_{uB\phi}^{33}$, the expected sensitivity at the ILC 
shall largely surpass the current and potential LHC limits \cite{Baur:2004uw,Aad:2012ux}.

\section{Electron beam polarization}

The use of electron beam polarization provides a competitive advantage to the ILC, not acessible at the LHC.
In particular, the beam polarization allows to separate the $d_j^Z$ and $d_j^\gamma$ couplings, or in other words, 
the different anomalous contributions to the $Z$ and $\gamma$ exchange in the $s$-channel, 
due to the different dependencies in different polarized beam scenarios.

\begin{figure}[ht]
\begin{center}
\begin{tabular}{cc}
\includegraphics[width=6.5cm]{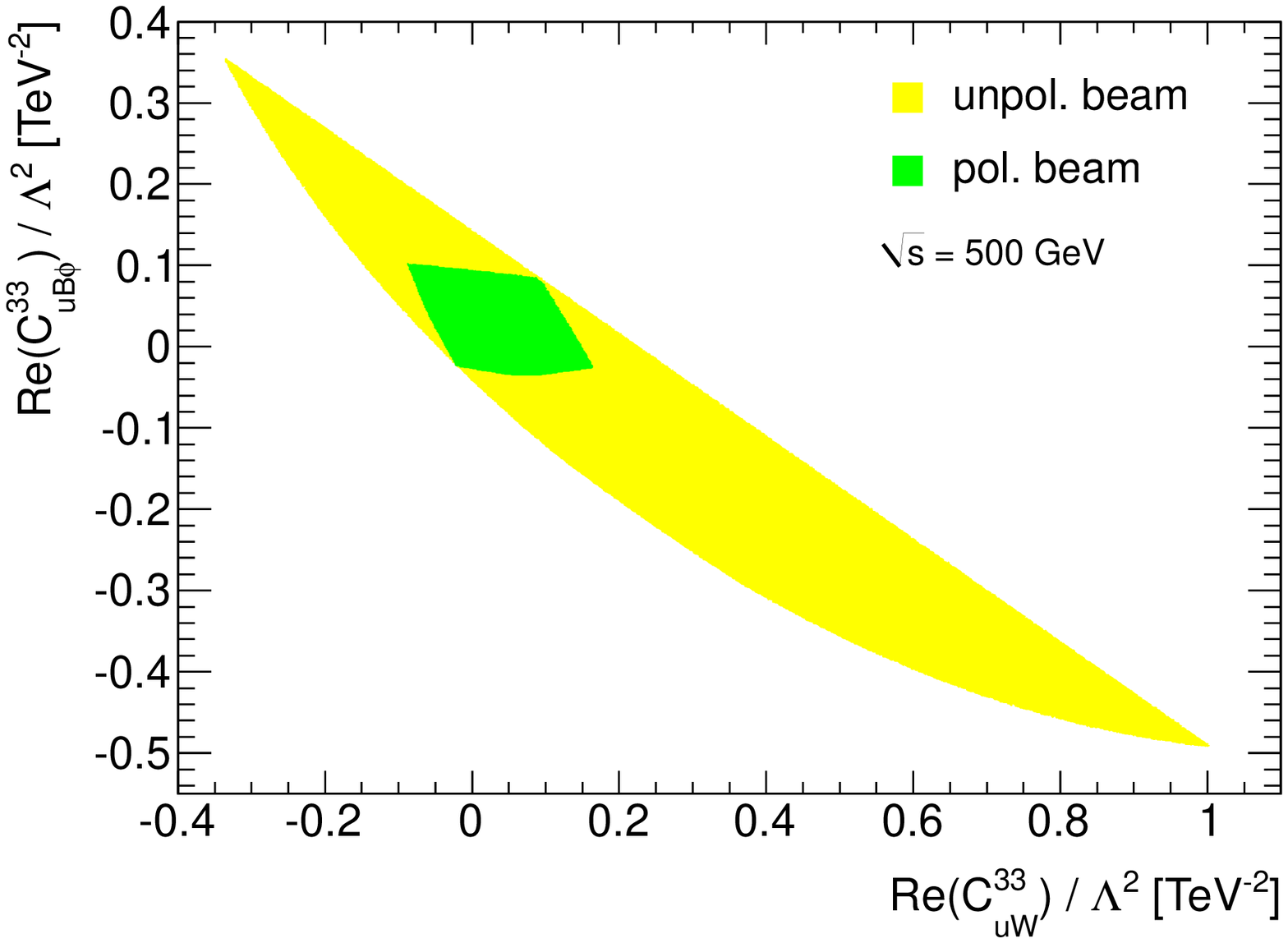} & \includegraphics[width=6.5cm]{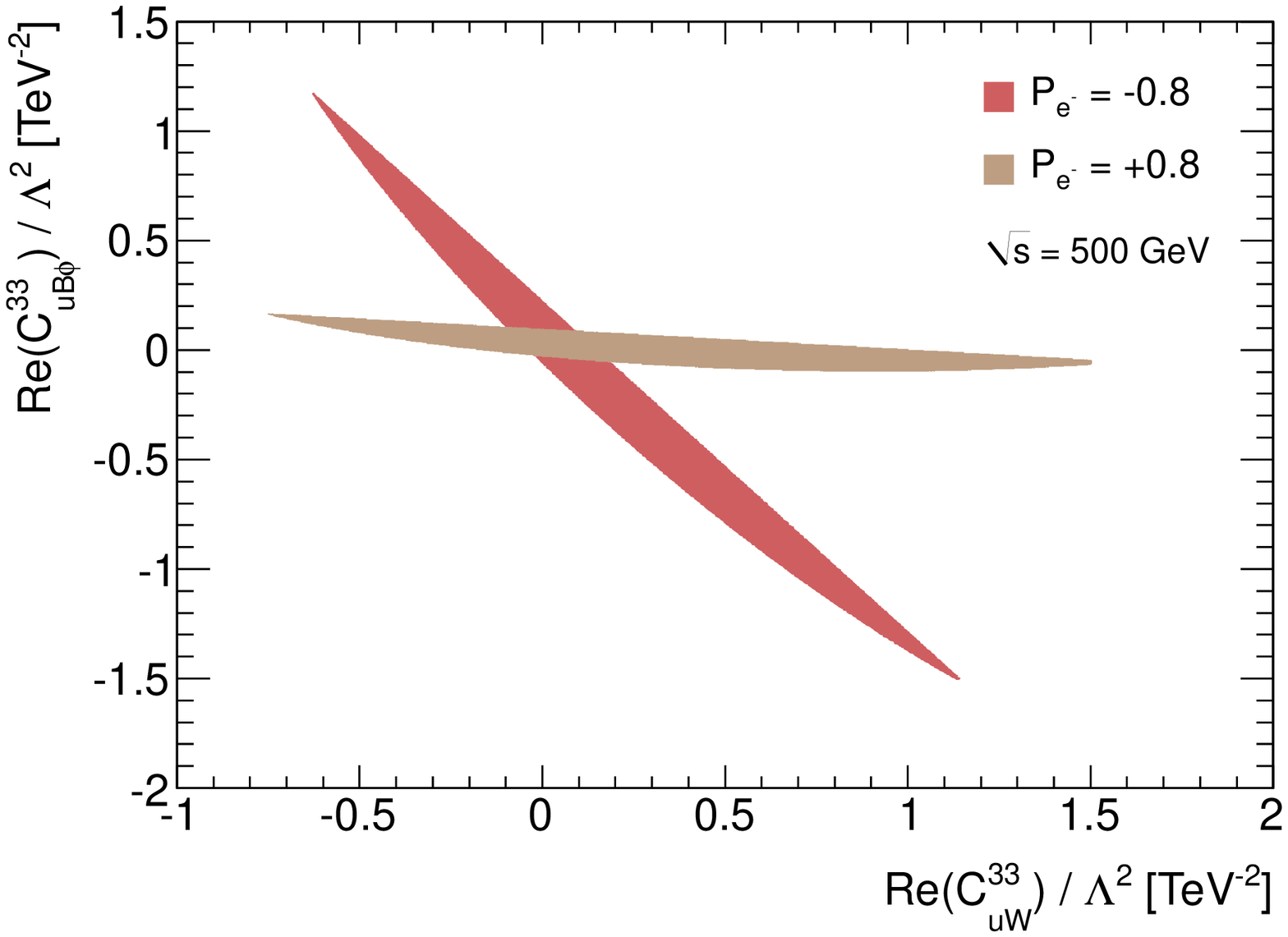}
\end{tabular}
\caption{Left: combined limits on $C_{uW}^{33}$ and $C_{uB\phi}^{33}$ for the cases of no beam polarization and electron beam polarization. 
Right: complementarity of the measurements for $P_{e^-} = 0.8$ and $P_{e^-} = - 0.8$.}
\label{fig:pol}
\end{center}
\end{figure}

In this preliminary study, two polarized beam scenarios are considered: $80\%$ right-handed ($P_{e^-} = 0.8$)
and $80\%$ left-handed ($P_{e^-} = -0.8$).
Since the possibility of using the polarization of the positron beam at a considerable value is still uncertain, its use is not taken into account here.
In Figure \ref{fig:pol}, the estimated allowed regions on $\RE \, C_{uW}^{33}$ and $\RE \, C_{uB\phi}^{33}$ are presented with 
and without the use of beam polarization (left plot), and for each individual polarized beam scenario (right plot).
In the left plot, the yellow region corresponds to the unpolarized case, where the measurements of both coefficients are anti-correlated, while 
the green region is much smaller due to the use of electron polarization.
On the right plot, the orthogonality of the two regions shows the complementary of the left- and right-handed beams, 
which allows to disentangle $C_{uW}^{33}$ and $C_{uB\phi}^{33}$.
It is, therefore, clear, how the use of different polarizations allows to constrain and disentangle these coefficients.
These results are obtained assuming the rest of operator coefficients are zero.

\section{Center-of-mass energy upgrade to 1 TeV}

\begin{figure}[t]
\begin{center}
\includegraphics[width=9.5cm]{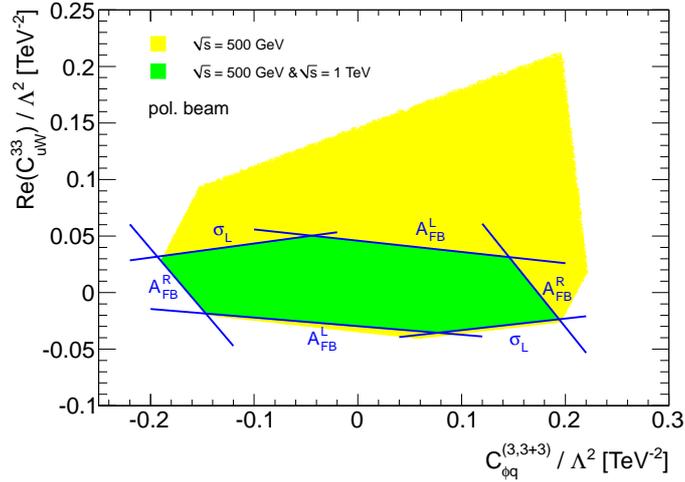}
\caption{Combined limits on $C_{\phi q}^{(3,3+3)}$ and $C_{uW}^{33}$ for a CM energy of 500 GeV and also with 1 TeV.}
\label{fig:E1000}
\end{center}
\end{figure}

Not only the use of electron beam polarization is useful in disentangling different operator contributions. 
The combination of measurements at different CM energies can also allow to separate the the vector and tensor contributions because the CM energy dependence is different.
Therefore, a possible upgrade to a CM energy of 1 TeV is crucial to distinguish $\gamma^\mu$ and $\sigma^{\mu \nu}$ couplings.
In Figure \ref{fig:E1000}, the allowed regions are shown for $C_{\phi q}^{(3,3+3)}$ and $C_{uW}^{33}$ coefficients, using polarized beams at 500 GeV (yellow), 
and using the combination of measurements at 500 GeV and 1 TeV (green). The total cross-sections and FB asymmetries for the different polarized beam scenarios were used 
to constrain these regions, and the blue lines around the green region correspond to the constraints caused by each observable\footnote{Note 
that these limits do not appear from a global fit but by requiring a $1\sigma$ agreement of the different observables considered.}.
Once more, becomes clear how the use of different observables in distinct experimental conditions can help to 
constrain the limits on the anomalous couplings and separate different contributions.

\section{Conclusions}

The effect of new physics contributions to the $t\bar{t}$ production at the ILC was estimated in this study, and compared with the LHC within an effective operator framework.
Even though the results at the LHC are already excellent, the sensitivity to the dimension-six effective operators coefficients is expected to be much better at the ILC than in LHC processes, 
such as top quark decays \cite{AguilarSaavedra:2007rs} and neutral current processes~\cite{Baur:2004uw,Baur:2005wi}.
These results can be further improved by considering observables in top quark decays at the ILC \cite{AguilarSaavedra:2012xe,Kane:1991bg,AguilarSaavedra:2006fy,AguilarSaavedra:2010nx}, 
and shall be taken into account in the future.
Finally, the combination of measurements with different electron beam polarizations at 500 GeV and 1 TeV allow to disentangle different contributions, and make the 
top quark studies a physics case for the use of polarizations, and for a possible upgrade to 1 TeV at the ILC.

\end{document}